\documentclass{mn2e}
\usepackage{epsfig}
\begin{document}
\title[IRAS 09111-1007]{Resolving $\bf IRAS~09111-1007$ at 350 microns: a different path to ULIRG formation?}
\author[Khan et al.]
{Sophia A. Khan$^{1,2}$, Dominic J. Benford$^2$, David L. Clements$^1$, 
S. Harvey Moseley$^2$,
\newauthor
Richard A. Shafer$^2$, Timothy J. Sumner$^1$ \\
$^{1}$Imperial College London, Blackett Laboratory, Prince Consort Road, London SW7 2AZ, UK\\
$^{2}$Infrared Astrophysics Branch, Code 685, NASA/GSFC, Greenbelt, MD 20771, USA\\
}
\maketitle

\begin{abstract}
We have resolved the ultraluminous infrared galaxy (ULIRG), $\rm
IRAS~09111-1007$, with the new 350$\,\mu$m-optimised Second Generation
Submillimeter High Angular Resolution Camera (SHARC~II) and present
the first submillimetre fluxes and images for the system.  $\rm
IRAS~09111-1007$ comprises two interacting luminous infrared galaxies
(LIRGs) with a projected nuclear separation of 39\,h$_{71}^{-1}\,$kpc.  The Western
galaxy is roughly four times more luminous in the submillimetre than its
Eastern counterpart.  It is an extremely bright LIRG with an AGN.  The
classification of the Eastern source is uncertain: it could be a
Seyfert 2 galaxy or a LINER.  We highlight $\rm IRAS~09111-1007$ as a
system that necessitates further study: a double AGN ULIRG whose
molecular gas content differs from other widely separated pairs and
whose ULIRG phase might not be explained by current multiple merger and/or
final stage ULIRG scenarios.

\end{abstract}

\section{Introduction}

Amongst the first results of extragalactic mid-IR astronomy was the
discovery of a small number of galaxies that emit the bulk of their
bolometric luminosity in the infrared (Low \& Kleinmann 1968, Kleinmann
\& Low 1970a,b).  The InfraRed Astronomical Satellite, IRAS, detected
large numbers of these ultraluminous infrared galaxies (ULIRGs)
(Soifer et al. 1984, Joseph \& Wright 1985, and Soifer, Neugebauer \&
Houck 1987)
with quasar-like luminosities of $\rm L_{IR}~(8$--$\rm 1000\,\mu m)~>
10^{12}\,L_\odot$.  There is still debate as to the nature of the far-IR
power source in these galaxies: is the immense thermal energy driven
by a dominant starburst, a dominant AGN or some combination of the
two?  These low-redshift IRAS-selected
ULIRGs are expected to be the counterparts to the high
redshift ($z >1$) SCUBA sources (see, e.g., Smail et al. 1998, Blain
et al. 2002, Webb et al. 2003, Chapman et al. 2003).

Most ULIRG systems have been shown to be disturbed, interacting or
merging in some way when the separation of nuclei is less than 10kpc
(Sanders et al. 1988, Clements et al. 1996, Murphy et al. 1996, 
Farrah et al. 2001).  The nature of widely separated ULIRG systems is less clear
(Dinh-V-Trung et al. 2001, Meusinger et al.  2001): are the components
of the ULIRG (supposedly the end phase of the galactic interaction)
beginning another merger or is the ULIRG a result of a multiple merger
event (Borne et al. 2000)?  This latter scenario is possible in widely
separated ULIRGs with resolved double nuclei but might not apply to the
ULIRG system $\rm IRAS~09111-1007$, which consists of two widely
spaced but interacting luminous infrared galaxies (LIRGs), each with 
a single nucleus.  The two LIRGs have a projected separation of 39\,h$_{71}^{-1}\,$kpc and a 
velocity difference of $\rm 425\,km\,s^{-1}$ (Duc, Mirabel \& Maza
1997).  In this letter we present 350$\,\mu$m
resolved images and fluxes for $\rm IRAS~09111-1007$.  We model the
far-IR dust emission to constrain the nature of the interaction.

\section{Observations and Data Reduction}

\begin{table*}
    \begin{tabular}{lcccccc}
    Name & Coordinates & 350$\mu$m Flux  &
      log($\rm L_{FIR:SED}$) &
      log($\rm L_{FIR:IRAS}$) & log($\rm L_{IR}$)& SFR \\
         & (J2000) & [Jy]  & [{L}$_\odot$]& [{L}$_\odot$]  & [{L}$_\odot$] &
[M$_\odot$yr$^{-1}$] \\ \hline
$\rm IRAS~09111-1007$     & ~ & ~ & 11.91  & 11.86 & 12.09 & 81\\ \hline
$\rm IRAS~09111-1007$W & 09 13 36.4 -10 19 31.8 & 0.85$\pm$0.13 &
11.80  & 11.75 & 11.98 & 63\\ \hline
$\rm IRAS~09111-1007$E & 09 13 38.8 -10 19 21.5 & 0.23$\pm$0.04 &
11.25 & 11.20 & 11.43 & 18 \\ \hline
\end{tabular}
   \caption{Coordinates, 350$\mu$m fluxes, luminosities and star
formation rates for the system and both components of $\rm IRAS~09111-1007$ (IRAS
fluxes from Surace, Sanders \& Mazzarella 2004).  The FIR
($40-500\,\mu$m) and IR ($8-1000\,\mu$m) luminosities are derived as follows: 
   $\rm L_{FIR:SED}$ is computed from best-fitting single 
   temperature SED, 
   $\rm L_{FIR:IRAS}$ and $\rm L_{IR}$ are calculated using the
standard relations from Fullmer \&
Lonsdale (1989) and Sanders \& Mirabel (1996) respectively.  Note that the pair 
positions given in both NED and Surace, Sanders \& Mazzarella (2004) are swapped.}  
\label{tab:fluxes}
\end{table*}

The data were taken using the Second Generation Submillimeter High
Angular Resolution Camera (SHARC~II) at the Caltech Submillimeter
Observatory on Mauna Kea, Hawai'i, in January and March 2004.  SHARC~II is a
350$\,\mu$m-optimised camera (Dowell et al.  2003) built around a
$12\times 32$ element close-packed bolometer array (Moseley et al.
2004).  It achieves a point-source sensitivity of $\rm \sim
1\,Jy~Hz^{-1/2}$ in good weather.  The 384 pixels of the SHARC~II
array image a region of around $1.0' \times 2.5'$.  Its filled
absorber array provides instantaneous imaging of the entire field of
view, sampled at roughly 2.5 pixels per nominal beam area.  The beam
profile was measured on known compact sources, and was verified to be
within five per cent of the diffraction-limited beamwidth of $8.5''$.

For these data the in-band zenith atmospheric opacity ($\tau_{350\,\mu\rm m}$) ranged
from 1.1 to 1.3, corresponding to a zenith transmission of around 30
per cent.  Our observations were centred on the
Eastern source of the pair, at position RA$=09^h13^m38.^s6$,
Dec$=-10^\circ19'20''$ (J2000).

In the submillimetre, emission from the atmosphere dominates the
signal read out by the bolometers.  In order to detect the faint
celestial sources it is necessary to remove this emission.  The sky
signal, however, is largely correlated between pixels.  At any moment
in time, every pair of pixels can be used as a `signal' beam and a
`reference' beam.  Additionally, the telescope is moved such that each
position on the sky can be viewed by many detectors at different
times.  This introduces a self-consistent self-calibration of all
pixels compared with all others.  A Lissajous scan pattern is used to
ensure that the area on the sky is well covered and has substantial
redundancy of observing.

The data were reduced using the standard CSO reduction software, CRUSH
(Kov\'acs, in preparation).  This software implements a self-consistent
least-squares algorithm to solve for the celestial emission, taking
into account instrumental and atmospheric contributions to the signal.
All observations were taken using the Dish Surface Optimisation System
(Leong 2003), which corrects for the primary mirror deformation as a
function of zenith angle, to improve the telescope efficiency and
the pointing.

The skymap is calibrated with a point spread function based on all
point source observations (Callisto, Ceres \& Arp 220) through the
observing period at similar elevations.  An oversampled $\chi^{2}$ fit
is used to determine the position of the source and the flux per
beam.  The processed image is shown in Figure 1.

\begin{figure}
       \begin{center}
\epsfig{file=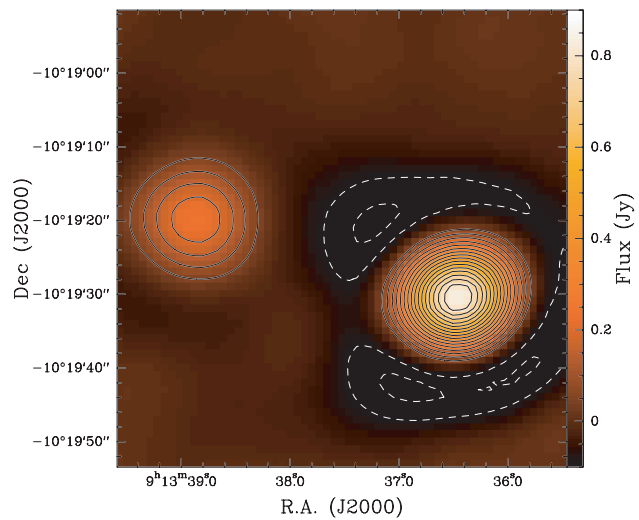,width=3.25in}
       \caption[SHARC Image]{350$\,\mu$m continuum emission map of
       $\rm IRAS~09111-1007$ taken with SHARC~II.  The contours are in
levels of $50\,$mJy/beam (dashed contours are negative and are an artifact of data reduction).}
        \label{fig:sharcmap}
        \end{center}
\end{figure}

\section{Results}

We are able to resolve the $\rm IRAS~09111-1007$ system and obtain
350$\,\mu$m fluxes and positions.  An absolute pointing offset of
$2.0''$ ($\Delta \alpha=1.0''$, $\Delta \delta=1.7''$) from the 2MASS
positions was attributed to a pointing error (due to the imprecise 
absolute pointing knowledge of the CSO) and removed from the SHARC~II 
image presented in this letter. 

The brighter LIRG, that which contributes $\approx${79} per cent of the
total system 350\,$\mu$m emission, is known as the `Western source' or
$\rm IRAS~09111-1007$W for the purposes of this letter.  The Eastern
LIRG is called the `Eastern source' or $\rm
IRAS~09111-1007$E (Murphy et al. 1996).  Together they form the
`ULIRG system'.  Figure 2 shows the resolved components of $\rm
IRAS~09111-1007$ imaged with the DSS, 2MASS and IRAS (HIRES processed;
Surace, Sanders \& Mazzarella 2004) respectively.  The 350$\,\mu$m
fluxes for each component are presented in Table 1.  The 
signal--to--noise in the detection is 89 for the 
Western source and 29 for the Eastern source.  The Western source 
is roughly four times more luminous in the submillimetre than its Eastern counterpart.

\begin{figure*}
       \begin{center}
\epsfig{file=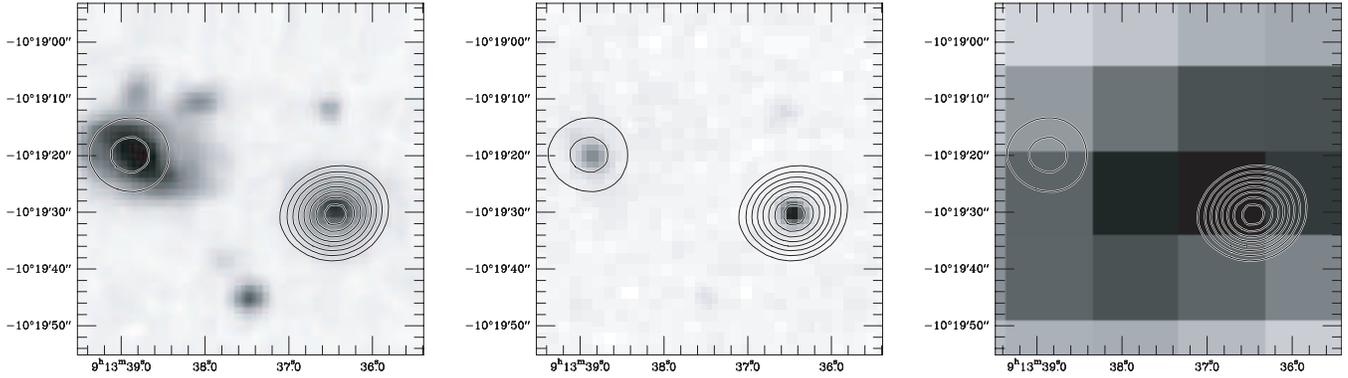,width=7in}
       \caption[Multiband Imaging]{DSS Optical, 2MASS K$_{s}$, and
       IRAS 60$\,\mu$m images of $\rm IRAS~09111-1007$, with
350$\,\mu$m contours overlaid on each (in levels of 100\,mJy/beam).}
        \label{fig:multi}
        \end{center}
\end{figure*}

\section{SED Modelling}

\subsection{Dust Temperature Blackbody Fitting}

\begin{figure}
     \begin{center}
    \epsfig{file=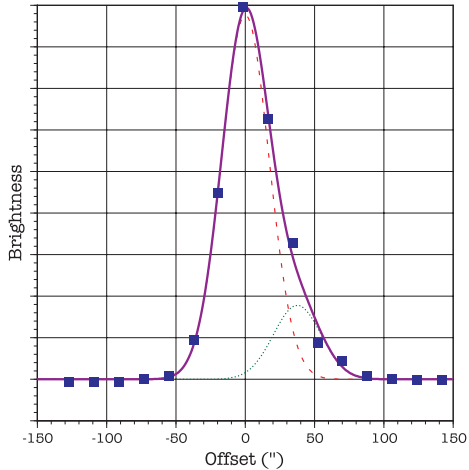,width=2.4in}
        \caption[Slice]{Fit to the spatial variation in intensity for
the HIRES-processed IRAS $60\,\mu$m image, cut through both galaxies.}
      \label{fig:slice}
      \end{center}
\end{figure}

A modified blackbody is used to model the total dust emission of the
$\rm IRAS~09111-1007$ system, specifically:
\begin{equation}
F_\nu = (M_{dust}/D^{2}) \kappa(\lambda) B_\nu(\lambda, T_{dust})
\end{equation}
where $B_{\nu}$ is the Planck function, $\kappa$ the mass absorption
coefficient of the dust ($\kappa(\lambda) \propto \lambda^{-\beta}$, see Dwek 2004), 
$\rm T_{dust}$ and $\rm M_{dust}$ the
equilibrium dust temperature and mass respectively, and $D$ the
distance of the galaxy.  For the distance we use WMAP cosmology ($\rm H_0 = 71\,km\,s^{-1}\, Mpc^{-1}$,
$\Omega_{m}=0.27$, and $\Omega_{\Lambda}=0.73$ (Bennett et al. 2003)).

In order to better constrain the SED of each component, we determined
the ratio of the fluxes of the two components at 60$\,\mu$m.  Beginning
with the IRAS 60$\,\mu$m HIRES-processed image of Surace et al. 
(2004), we sliced along the axis of the known 350$\,\mu$m
sources.  A pair of Gaussian intensity functions were fitted to the
measured flux along this slice, constraining them to have the same
width and a fixed spacing as determined at 350$\,\mu$m (see
Figure 3).  The remaining four free parameters (one position,
the width, and the two intensities) are then well-constrained, with
the flux ratio being $4.9\pm 0.8$ (the total system flux as measured
by IRAS is still valid, since the IRAS fluxes are derived with an
aperture large compared to the source separation).  The ratio derived
from the $350\,\mu$m measurement is $3.7\pm1.1$.  Alternatively, the
fraction of the flux from the brighter LIRG is $83\pm3$ per cent at
$60\,\mu$m and $79\pm16$ per cent at 350$\,\mu$m, an entirely consistent
measurement.  The ratio of 79 per cent is accurate enough for the fits
that follow.

$\chi^{2}$ minimization determined the best-fitting emissivity
index ($\beta$) and corresponding single fit dust temperature for the 
system using
the SHARC~II 350$\,\mu$m flux with the 60$\,\mu$m and 100$\,\mu$m IRAS 
fluxes derived in Surace et al. (2004) (the 100$\,\mu$m flux was used 
as the
normalization).  The best-fitting was $\beta$=1.9$\pm$0.1, T=31$\pm1\,$K, and is shown in
Figure 4.  We also model both the Western and Eastern sources assuming
that the IRAS fluxes are distributed in the same way as the
350$\,\mu$m emission. Dust masses are calculated assuming the 100$\,\mu$m mass absorption coefficient, 
$\kappa_{100}$, is 40$\rm \,cm^2\,g^{-1}$ (Draine \& Lee 1984).

\begin{table}
      \begin{tabular}{lcccc}
Model & Source         & $\beta$     & $\rm T_{dust}$   & $\rm M_{dust}$\\
   & & & [K] & [10$^{6}$M$_\odot$] \\
\hline \hline
Single Temperature & System & $1.9\pm0.1$ & $31\pm1$ & 220$^{+30}_{-20}$\\ 
 & Western source &  $1.9\pm0.1$ & $31\pm1$ & 170$^{+30}_{-20}$\\ 
 & Eastern source & $1.9\pm0.1$ & $31\pm1$  & 50$^{+10}_{-10}$ \\ \hline
Two Temperature & System & 2.0 & $32\pm2$  & 300$^{+190}_{-110}$ \\ 
$\rm M_{cold}/M_{warm}$=1 & & (fixed) & $26\pm5$ & \\ \hline
Two Temperature & System & 2.0 & $36\pm3$ & 280$^{+70}_{-50}$ \\
$\rm M_{cold}/M_{warm}$=11 & & (fixed) & $29\pm1$ & \\ \hline
Single Temperature & System  & n/a     & $30\pm1$  & 250$^{+30}_{-30}$\\ 
DL Dust Model & Western source & n/a     & $30\pm1$  & 200$^{+30}_{-30}$\\ 
& Eastern source & n/a     & $30\pm1$  & 60$^{+10}_{-10}$ \\ \hline

      \end{tabular}
   \caption{Best-fitting SED parameters: $\beta$, temperature and dust mass for the
$\rm IRAS~09111-1007$ system and components, with corresponding estimated 
1$\sigma$ uncertainties.}
    \label{tab:sed}
\end{table}

With only two independent colours we cannot constrain the cold and 
warm dust masses and simultaneously fit the dust temperatures in a two temperature 
model.  Using $\beta$=2 (as in Dunne \& Eales 2001), if we assume equal amounts of cold and 
warm dust in the system we get best-fitting temperatures of $\rm 
T_{cold}=26\pm5$\,K and 
$\rm T_{warm}=32\pm2\,$K.  If we adopt a dust 
mass distribution typical of the highest 
luminosity galaxies from the SCUBA Local Universe Galaxy Survey
(SLUGS), as given in Dunne \& Eales (2001), we get a ratio $\rm 
M_{cold}/M_{warm}$=11.  The best-fitting values are then 
$\rm T_{cold}=29\pm1\,K, T_{warm}=36\pm3$\,K.

As an alternative to these approaches, we also used a more physically-based 
dust grain model (the single temperature DL dust model -- Draine \& Lee 1984,
Laor \& Draine 1993), an equal mixture of silicates and graphites 
with $\kappa_{100}$ of 31 and 54$\rm 
\,cm^2\,g^{-1}$ for the silicate and graphite grains respectively.  In this
case, the best-fitting temperature of the system was $30\pm1\,$K.   

The dust temperature of $\sim$ 31\,K is within 1.5$\sigma$ of the average temperature of
$38\pm6\,$K from the previous 350$\,\mu$m LIRG study by Benford 
(1999), with the first SHARC camera (although that survey did not include any
widely separated LIRG systems).  The result that the 60$\,\mu$m dust is distributed 
in very similar ratios to the 350$\,\mu$m dust (Figure 3) is responsible for the 
identical SED temperatures for both components.

\subsection{Ratio of Molecular Gas To Dust}

The molecular gas mass in the Western source is 
$\rm 2.3 \times 10^{10}\,M_\odot$ (Mirabel et al. 1990).  The Western 
source's dust mass is derived from the best-fitting SED to give a 
molecular gas--to--dust ratio of 140 and 120 for the single 
temperature and DL dust models respectively.  In the absence of 21\,cm H{\sc i} 
observations this value is a lower limit on the total 
(molecular + atomic) gas--to--dust ratio.

The 8--1000\,$\mu$m luminosity ($\rm L_{IR}$, see Table 1) 
gives a value for the Western source IR luminosity--to--H$_{2}$ mass ratio 
($\rm L_{IR}$/M(H$_{2}$) -- the star formation 
efficiency) of 42$\rm\,L_\odot\,M_\odot^{-1}$.  The molecular 
gas--to--dust and IR luminosity--to--H$_{2}$ mass ratios
are consistent with values for the highest luminosity SLUGS galaxies (Dunne et al. 2000).

Dinh-V-Trung et al. (2001) studied six widely separated ($>20\,$kpc)
ULIRG systems in the complete 1\,Jy sample of Kim \& Sanders (1998).  
In their sample, the molecular gas was concentrated 
in the dominant source of the far-IR emission.  Although the Western 
source of $\rm IRAS~09111-1007$ is gas-rich, the Eastern source, by 
virtue of the amount of dust, is unlikely to be gas-poor.  This 
would make this system different from their wide-pair ULIRG sample.

\begin{figure}
     \begin{center}
    \epsfig{file=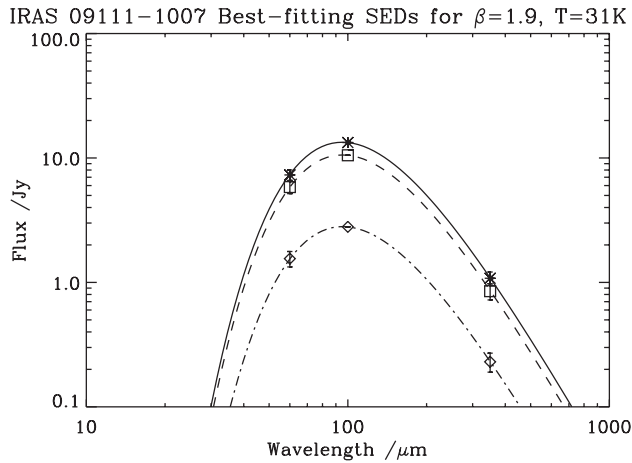,width=3.25in}
        \caption[SED]{Single Temperature Best-Fitting SED for
the Western source (dashed), Eastern source (dot-dashed), and
the $\rm IRAS~09111-1007$ system (solid: $\beta$=1.9,
T=31$\,$K).}
      \label{fig:sed}
      \end{center}
\end{figure}

\subsection{Star Formation Rate}

The system and component luminosities (see Table 1) show that the two 
components form a ULIRG system, though neither is a ULIRG 
individually.  The star formation rate in Table 1 uses the SED derived 
FIR luminosities and the relation of Thronson $\&$ Telesco (1986):

\begin{equation}
     \rm  SFR\sim \Psi
10^{-10}\,\Bigl(\frac{L_{FIR}}{L_{\odot}}\Bigr)\,M_{\odot}\,yr^{-1}
      \end{equation}
assuming $\Psi$ = 1 (typical values of $\Psi$ are 0.8--2.1).

\section{Discussion}

\subsection{Source Characterisation}

The optical image from the Digitized Sky Survey (Figure 2, left) shows
the disturbed morphology of the Eastern source.  In the 2MASS
K$_{s}$ band image (Figure 2, centre) the Western source is
slightly more luminous than the Eastern, a trait even more prominent
in the submillimetre (Figure 1).  Merger models such as Barnes $\&$
Hernquist (1991) predict gas and dust to be concentrated in the galaxy
centre as the ULIRG interaction condenses large amounts of the ISM
into the nuclear region.

Line ratios from Duc et al. (1997) classify the Western source as a
Seyfert 2.  Dudley (1999) found prominent polycyclic aromatic
hydrocarbon features in the 8--13$\,\mu$m dust emission spectra --
indicative of a starburst.  The Eastern source is either a Seyfert 2 or
LINER galaxy (Duc et al. 1997), while Gon\c{c}alves, V\'{e}ron-Cetty
\& V\'{e}ron (1999) also find evidence for a starburst.  Neither of
the two sources in the $\rm IRAS~09111-1007$ system were detected in
the ROSAT All-Sky Survey.  A non-detection in the ROSAT band does not necessarily mean
that a source is intrinsically weak since the soft X-ray
band is sensitive to X-ray absorption, which is common in AGN.

\subsection{Merging Stage}

Without additional submillimetre data we are unable to constrain the 
relative temperatures and masses of the cold and warm dust components.  Whether 
the dust temperature of the system
is related to the stage of merging is not clear.  Mazzarella, Bothun \& 
Boroson (1991) find an increase in warm dust temperature with merging 
stage, although Klaas et al. (2001) argue the cold dust temperature 
would increase as well.  The wide separation of the pair would suggest that $\rm IRAS~09111-1007$ is at
the beginning of a merger, a notion supported by the value 
of the IR luminosity--to--H$_{2}$ mass ratio, which falls within a region of $\rm L_{IR}/M(H_2)$ 
vs $\rm L_{IR}$ space that is common for early merging systems (Sanders, Scoville \& 
Soifer 1991).  

\subsection{Widely Spaced ULIRG Pairs}

Unlike the wide pair sample of Xu \& Sulentic (1991) {\it both}
components are enhanced in the far-IR.  With a velocity difference
of $\rm 425\,km\,s^{-1}$ (Duc et al. 1997) and a projected separation
of 39\,h$_{71}^{-1}\,$kpc, Monte Carlo simulations give the probability of the pair being
bound as 0.88 (Schweizer 1987).  

Although widely separated ULIRG pairs are not uncommon the nature of
their interaction is still uncertain.  Murphy et al. (1996)
postulated the presence of a third nucleus in widely
separated pairs, though no double nucleus has been detected in either
component of $\rm IRAS~09111-1007$.  The galaxies were shown to be unresolved at 
$0.''5$ resolution in a $6\,$cm search by Crawford et al. (1996).  However
the non-detection of a double nucleus cannot rule out a multiple
merger since the time-scale of nuclei
coalescence is short (Surace, Sanders \& Evans 2000, Meusinger et al.
2001).  In multiple approach merger models (e.g., Dubinski, Mihos \& Hernquist
1999) the merging process is a series of encounters
where bound components approach and separate.  A previous encounter may have triggered the
starburst/AGN in the system.  High resolution optical imaging (Borne 
et al. 2000) could decide between these scenarios by either detecting multiple 
nuclei or confirming single nuclei.  High resolution CO imaging would 
be needed to detect whether gas has been disturbed by a previous phase 
of the merging event (Mihos \& Hernquist 1996).

\section{Conclusions}

We have resolved the widely separated ULIRG system of $\rm
IRAS~09111-1007$ with the SHARC~II detector at 350$\,\mu$m.  This
system comprises two LIRGs with a projected separation of
39\,h$_{71}^{-1}\,$kpc, or around two optical diameters.  The Western
component dominates the far-IR flux at both 60$\,\mu$m and
$350\,\mu$m, carrying 79 per cent of the total system luminosity.
Although the luminosity of the system is large, our fluxes
suggest a dust temperature of $31\,$K for this system, with both
components at the same temperature to within the sensitivity of this
measurement.  
The wide separation and the value of the IR 
luminosity--to--H$_{2}$ mass ratio suggest that the pair are at an 
early stage of interaction.  But the high luminosity of the system 
($\rm L_{IR}=1.2\times 10^{12}$\,L$_\odot$) would be unusual for 
such a stage, unless the components had experienced a previous 
merger or interaction.  A high resolution optical search 
for multiple nuclei within each component is needed.  Their absence 
could indicate that the double AGN--LIRG system of $\rm IRAS~09111-1007$, 
and perhaps other widely spaced ULIRG pairs, 
might be unexplained by current theories of ULIRG formation and evolution.

\section{ACKNOWLEDGMENTS}

This research has made use of the NASA/IPAC Extragalactic Data base
(NED), which is operated by the Jet Propulsion Laboratory under
contract with NASA.  The
optical image is taken from photographic data obtained using the UK
Schmidt Telescope.  The Caltech Submillimeter Observatory is supported by
NSF contract AST-0229008.

We thank the anonymous referee for their insightful 
and careful comments.  We are very grateful to Pierre-Alain Duc for providing
clarification and additional spectroscopic data for the system, to
Jason Surace for providing his HIRES-processed IRAS images, to Eli
Dwek for SED discussion and the DL dust model, and to Rick Arendt for
invaluable IDL data analysis advice.  We extend our thanks to Tom 
Phillips and the CSO for our observing time and
support during our runs, Hiroshige Yoshida for undertaking heterodyne
observations on our behalf, and to Darren Dowell, Attila Kov\'acs, 
and Colin Borys for observation and instrument assistance and
support with data analysis.  

\section{References}

Barnes J.E., Hernquist L.E., 1991, ApJ, 370L, 65\\
Benford D.J., 1999, PhD thesis, Caltech\\
Bennett C.L., et al., 2003, ApJS, 148, 1\\
Blain A.W., Smail I., Ivison R.J., Kneib J.-P., Frayer D.T., 2002, 
PhR, 369, 111\\
Borne K.D., et al., 2000, ApJ, 529L, 77\\
Chapman S.C., Blain A.W, Ivison R.J., Smail I.R., 2003, Nature,
422, 695\\
Clements D.L., Sutherland W.J., McMahon R.G., Saunders W., 1996, MNRAS,
279, 477\\
Crawford, T., Marr, J., Partridge, B., Strauss, M.A., 1996, ApJ, 460, 
225\\
Dinh-V-Trung, Lo K.Y., Kim D.-C., Gao Y., Gruendl R.A., 2001,
ApJ, 556, 141\\
Dowell C.D., et al., 2003, SPIE, 4855, 73\\
Draine B.T., Lee H.M., 1984, ApJ, 285, 89\\
Dubinski J., Mihos J.C., Hernquist L., 1999, ApJ, 526, 607\\
Duc P.-A., Mirabel I.F., Maza J., 1997, A\&AS, 124, 533\\
Dudley C.C., 1999, MNRAS, 307, 553\\
Dunne L., Eales S., Edmunds M., Ivison R., Alexander P., Clements 
D.L., 2000, MNRAS, 315, 115\\
Dunne L., Eales S.A., 2001, MNRAS, 327, 697\\
Dwek E., 2004, ApJ, 607, 848\\
Farrah D., et al., 2001, MNRAS, 326, 1333\\
Fullmer L., Lonsdale C.J., 1989, JPL D-1932, Version 2, part no 3\\
Gon\c{c}alves A.C., V\'{e}ron-Cetty M.P., V\'{e}ron P., 1999,
A\&AS, 135, 437\\
Joseph R.D., Wright G.S., 1985, MNRAS, 214, 87\\
Kim D.-C., Sanders D.B., ApJS,  1998, 119, 41\\
Klaas U., et al., 2001, A\&A, 379, 823\\
Kleinmann D.E., Low F.J., 1970a, ApJ, 159L, 165\\
Kleinmann D.E., Low, F.J., 1970b, ApJ, 161L, 203\\
Laor A., Draine B.T., 1993, ApJ, 402, 441\\
Leong M., 2003, http://www.cso.caltech.edu/dsos/DSOSAMOSpaper.htm\\
Low J., Kleinmann D.E., 1968, AJ, 73, 868\\
Mazzarella J.M., Bothun G.D., Boroson T.A., 1991, ApJ, 101, 2034\\
Meusinger H., Stecklum B., Theis C., Brunzendorf J., 2001, A\&A,
379, 845\\
Mihos J.C., Hernquist L., 1996, ApJ, 464, 641\\
Mirabel I.F., Booth R.S., Johansson L.E.B., Garay G., Sanders D.B., 
1990, A\&A, 236, 327\\
Moseley S.H., Allen C.A., Benford D., Dowell C.D., Harper D.A.,
Phillips T.G., Silverberg R.F., Staguhn J., 2004, NIMPA, 520, 417\\
Murphy T.W.,Jr, Armus L., Matthews K., Soifer B.T., Mazzarella J.M.,
Shupe D.L., Strauss M.A., Neugebauer G., 1996, AJ, 111, 1025\\
Sanders D.B., Soifer B.T., Elias J.H., Madore B.F., Matthews K.,
Neugebauer G., Scoville N.Z., 1988, ApJ, 325, 74\\
Sanders D.B., Scoville N.Z., Soifer B.T., 1991, ApJ, 370, 158\\
Sanders D.B., Mirabel I.F., 1996, ARA\&A, 34, 749\\
Schweizer L.Y., 1987, ApJS, 64, 427\\
Smail I., Ivison R.J., Blain A.W., Kneib J.-P., 1998, ApJ, 507L,
21\\
Soifer B.T., et al., 1984, ApJ, 278L, 71\\
Soifer B.T., Neugebauer G., Houck J.R., 1987, ARA\&A, 25, 187\\
Surace J.A., Sanders D.B., Evans A.S., 2000, ApJ, 529, 170\\
Surace J.A., Sanders D.B., Mazzarella J.M., 2004, AJ, 127, 3235\\
Thronson H., Telesco C., 1986, ApJ, 311, 98\\
Webb T.M.A, Lilly S.J., Clements D.L., Eales S., Yun M., Brodwin M.,
Dunne L., Gear W.K., 2003, ApJ, 597, 680\\
Xu C., Sulentic J.W., 1991, ApJ, 374, 407\\

\end{document}